# Hadron Showers in a Digital Hadron Calorimeter

Burak Bilki[d], John Butler[b], Georgios Mavromanolakis[c,1], Ed May[a], Edwin Norbeck[d], José Repond[a], David Underwood[a], Lei Xia[a], Qingmin Zhang[a,2]

[a]*Argonne National Laboratory, 9700 S. Cass Avenue, Argonne, IL 60439, U.S.A.*
[b]*Boston University, 590 Commonwealth Avenue, Boston, MA 02215, U.S.A.*
[c]*Fermilab, P.O. Box 500, Batavia, IL 60510-0500, U.S.A.*
[d]*University of Iowa, Iowa City, IA 52242-1479, U.S.A.*

**Abstract.** A small prototype of a finely granulated digital hadron calorimeter with Resistive Plate Chambers as active elements was exposed to positive pions of 1 – 16 GeV energy from the Fermilab test beam. The event selection separates events with mostly non-interacting particles and events with hadronic showers which initiated in the front part of the calorimeter. The data are compared to a Monte Carlo simulation of the set-up. The paper concludes with predictions for the performance of an extended digital hadron calorimeter.

## INTRODUCTION

Particle Flow Algorithms (PFAs) attempt to measure all particles (originating from the interaction point of a typical colliding beam detector) in a jet individually, using the detector component providing the best momentum/energy resolution [1,2]. In this context and in preparation for the construction of a lager calorimeter module, a small prototype of a finely granulated hadron calorimeter (HCAL), using Resistive Plate Chambers (RPCs) as active elements, was assembled. The prototype featured 1 x 1 $cm^2$ readout pads and a total of 1536 channels in six layers, interleaved with absorber plates. The readout system applied a single threshold to each pad (corresponding to a 1-bit resolution), hence the designation of Digital Hadron Calorimeter (DHCAL). The stack was exposed to pions of the Fermilab test beam in the 1 – 16 GeV energy range. Measurements of the response functions are presented and compared to expectations from Monte Carlo simulations based on GEANT4 [3] and a standalone program (RPCSIM), modeling the response of RPCs. Based on the satisfactory agreement between test beam data and simulations, the same simulation tools are used to predict the performance of an extended DHCAL.

---

[1] Also affiliated with University of Cambridge, Cavendish Laboratory, Cambridge CB3 OHE, U.K. Now at CERN, Geneva, Switzerland.
[2] Also affiliated with Institute of High Energy Physics, Chinese Academy of Sciences, Beijing 100049, China and Graduate University of the Chinese Academy of Sciences, Beijing 100049,China.

This research was performed within the framework of the CALICE collaboration [4], which develops imaging calorimetry for the application of PFAs to the measurement of hadronic jets at a future lepton collider.

## DESCRIPTION OF THE CALORIMETER STACK

The calorimeter stack consisted of six chambers interleaved with absorber plates. Each absorber plate contained 16 mm thick steel and 4 mm thick copper, corresponding to approximately 1.2 radiation lengths. The overall depth of the calorimeter amounted to about 0.65 nuclear interaction lengths. The chambers measured 20 x 20 cm$^2$ in area and featured two glass plates. The thickness of the glass plates was 1.1 mm and the gas gap was maintained with fishing lines with a diameter of 1.2 mm.

The chambers were operated in avalanche mode with a high voltage setting of 6.3 kV. The gas consisted of a mixture of three components: R134A (94.5%), isobutane (5.0%) and sulfur-hexafluoride (0.5%) [5]. For additional details on the design and performance of the chambers, see [6,7].

The chambers were mounted on the absorber plates and these in turn were inserted into a hanging file structure. The gap between absorber plates was 13.4 mm, of which 8.3 mm were taken by the chambers and their readout boards.

The electronic readout system was optimized for the readout of large numbers of channels. In order to avoid an unnecessary complexity of the system, the charge resolution of individual pads was reduced to a single bit (digital readout). The readout system consisted of several parts: the pad-boards covering an area of 16 x 16 cm$^2$, the front-end board, the front-end Application Specific Integrated Circuits (the so-called DCAL chips), the data concentrator and data collector modules, and the timing and triggering module. For more details on the readout system see ref. [8].

Each layer contained 256 individual readout pads with an area of 1 x 1 cm$^2$. The entire stack had 1536 readout channels, of which only ten appeared to be dead and provided no signal. A photograph of the calorimeter stack in the test beam is shown in Fig. 1.

## TEST BEAM SETUP AND DATA COLLECTION

The stack was exposed to pions from the test beam at the Meson Test Beam Facility (MTBF) of Fermilab [9]. Pions were produced with an upstream target and were momentum selected in the range between 1 and 16 GeV/c. The beam came in spills of

3.5 second length every one minute and contained a mixture of positrons, muons, pions and protons.

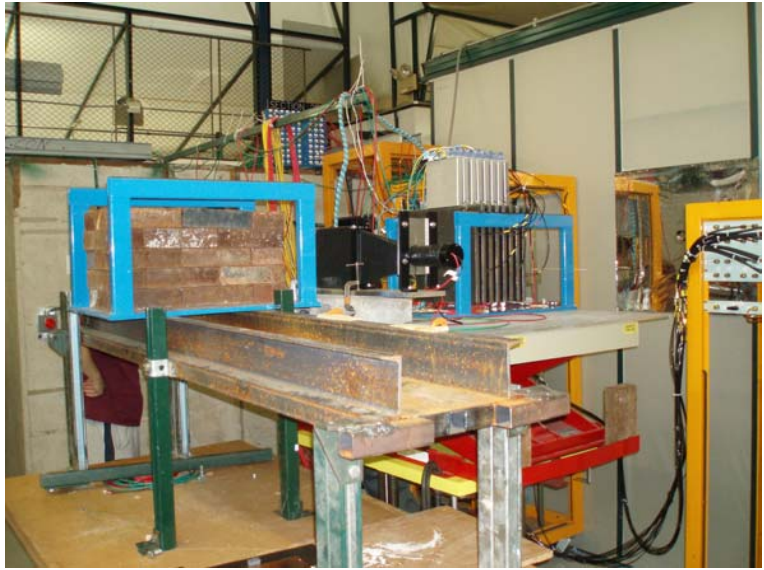

**Figure 1**. Photograph of the hanging file structure containing eight layers of which the first six were utilized for the present measurements. For one particular run, a stack of iron bricks was placed in front of the calorimeter, here seen to the left of the stack.

The readout of the stack was triggered by the coincidence of two large scintillator paddles, and was vetoed by two upstream Čerenkov counters. The scintillator paddles measured 19 x 19 cm$^2$ and were located approximately 2.0 and 0.5 meters upstream of the stack. The Čerenkov counters rejected positrons efficiently, but not entirely (see below). No attempt was made to identify and reject muons in the beam. Table I lists the number of triggers collected at each momentum setting together with the average beam intensity during a spill and the fraction of events without veto from the Čerenkov counters.

| Momentum [GeV/c] | Stack of iron bricks | Number of events | Beam intensity [Hz] | Fraction of events without veto from the Čerenkov counters[%] |
|---|---|---|---|---|
| 1 | No | 1378 | 547 | 6.0 |
| 2 | No | 5642 | 273 | 5.9 |
|   | Yes | 1068 | 80 | 57.3 |
| 4 | No | 5941 | 294 | 15.5 |
| 8 | No | 30657 | 230 | 24.6 |
| 16 | No | 29889 | 262 | 28.0 |

**Table I.** Summary of the pion runs.

At the front face of the stack the beam spot, for momenta between 4 and 16 GeV/c, was somewhat collimated with a sigma of approximately 2 cm, both horizontally and vertically. At the lowest two energies the beam spot appeared to cover the entire readout area of the chambers. Figure 2 shows various views/projections of an event with an 8 GeV/c pion undergoing an interaction between the $1^{st}$ and $2^{nd}$ layers.

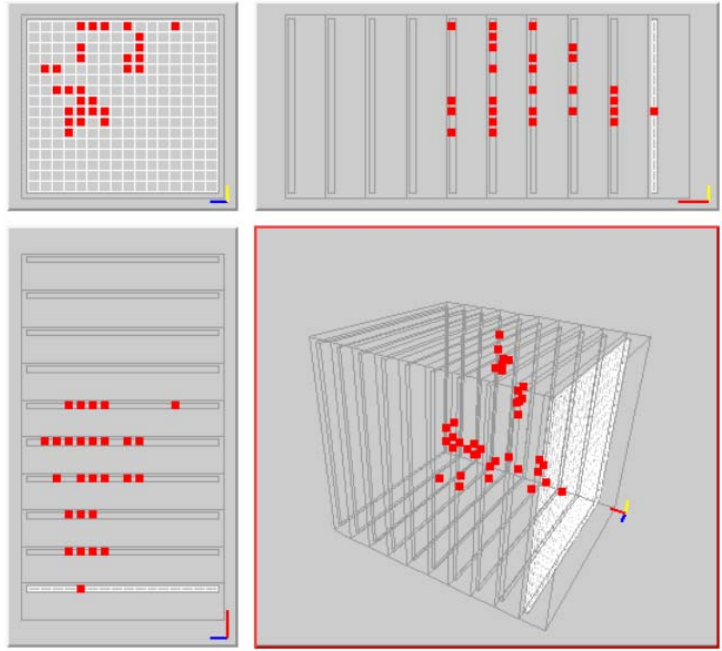

**Figure 2.** Event display of a pion induced shower with an interaction between the $1^{st}$ and $2^{nd}$ layers.

## MONTE CARLO SIMULATION AND CALIBRATION

The test beam set-up has been simulated with a Monte Carlo program based on the GEANT4 package [3] and a standalone program RPCSIM for the simulation of the response of RPCs. The GEANT4 simulation includes the relevant portions of the beam line, the trigger counters[3], and the details of the calorimeter stack. The program utilized the Linear Collider Physics List [9] and the range cut was left at its default value of 1.0 mm. Pion and positron data were generated at the various energies probed in the test beam and with their corresponding lateral beam profiles. The spatial coordinates of any energy deposition in the gas gap of an RPC was recorded for further analysis. In the following these energy depositions are named *points*.

Using the generated *points*, the RPCSIM program produced the corresponding hit patterns in each layer of the stack. For each *point* a charge was generated according to the induced charge distribution, as measured with an analog readout of the chambers

---

[3] The 16 GeV data showed an unusually large number of early showers. This data was taken before the data at other energies and before a re-arrangement and simplification of the beam line. Additional material (1/4 $X_0$ of iron) was introduced into the simulation of the beam line in order to obtain a satisfactory simulation of the measured longitudinal shower profile.

[6]. The details of the hit pattern generation was controlled by four parameters: **a** (the slope of the exponential decrease of the induced charge as function of lateral distance to the *point*), **T** (the threshold charge for registering a hit on a given pad), $Q_0$ (an adjustment to the measured charge distribution induced in the layer of pads), and $d_{cut}$ (a lateral distance in the gas gap within which there can be only one avalanche, independent of the number of *points*). The first three parameters were adjusted by comparison of the response of measured and simulated muon events in the calorimeter. The fourth parameter, $d_{cut}$, was determined with 4 GeV positron data and their comparison with simulated events. For additional details see [11]. In the following these parameters remained at their default values.

No attempt was made to simulate possible inefficiencies of the chambers [12] due to the high particle flux in pion and positron induced showers. However, for some momentum settings the overall response in the simulation had to be adjusted (up to -18%) to take into account these rate effects.

The Monte Carlo generated events were formatted the same way as the test beam data and were analyzed by identical offline programs.

## EVENT SELECTION

The event selection insured the high purity of the pion/muon data, while rejecting multi-particle events. The selection criteria are described in the following:

a) *Requirement of at least three layers with hits.* This cut removed contamination from accidental triggers.
b) *Requirement of exactly one cluster of hits in the first layer.* This cut effectively removed events with more than one particle entering the calorimeter or with showers which had initiated upstream of the calorimeter stack. Clusters of hits were reconstructed as aggregates of cells with at least one side in common between two cells.
c) *Requirement of no more than four hits in the first layer.* This cut also removed events with electromagnetic or hadronic showers which had initiated upstream of the calorimeter.
d) *Fiducial cut on the position of the cluster in the first layer.* In order to contain the showers laterally, the cluster in the first layer was required to be at least 3 cm from the edge of the readout area of the chambers. To reduce efficiency losses due to rate effects (see below), for the 8 GeV data an additional fiducial cut excluded an area of 2.5 x 2.5 $cm^2$ at the center of the chambers.
e) *MIP and shower selection:* Using the response of the second layer of the stack the data sample was split into two parts: 1) Requiring at most four hits in this layer a sample enriched with non-interacting particles was obtained. In the following this sample is labeled the 'MIP selection', and 2) Requiring at least five hits in this layer a sample enriched in showers was obtained. Using the second layer ensured that these showers started early in the stack. In the following this sample is named the 'shower selection'.

## ANALYSIS OF THE MUON DATA

In order to suppress the pion content in the beam and to obtain a sample of broadband muons traversing the calorimeter, a stack of iron bricks was piled up in front of the calorimeter. The stack measured 50 cm in depth, corresponding to about 3 nuclear interaction lengths, and covered the entire beam spot, see Fig. 1. The momentum selection for the secondary beam was set to 2 GeV/c. The muons lost on average 600 MeV in the iron stack, but retained enough energy to traverse the six layers of the calorimeter.

Applying selection criteria *a) – d)* provided a clean sample of muons, to calibrate the individual layers of the calorimeter. As a function of layer number, Fig. 3 shows the efficiency $\varepsilon_i$ (calculated as the ratio of events with at least one hit in layer i to all selected events), the pad multiplicity $\mu_i$ (calculated as the average number of pads firing in layer i when at least one hit is recorded in that layer) and the product of the two, $\varepsilon_i \mu_i$. Due to the requirement of exactly one cluster in the first layer (selection criterion *b)*) the efficiency of the first layer could not be measured by this method.

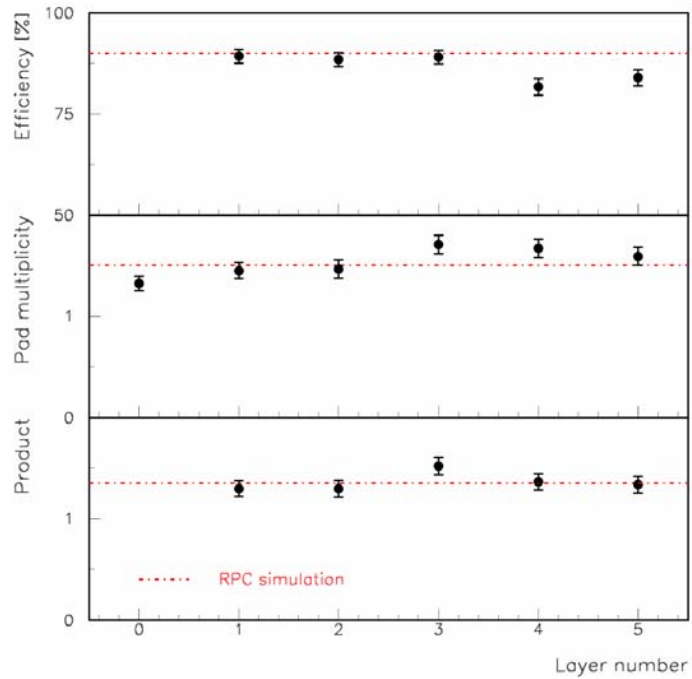

**Figure 3.** Average MIP detection efficiency (top), pad multiplicity (middle) and their product (bottom) for each layer of the stack. The dashed red lines indicate the values used in the simulation of the RPC response.

The measurements are compared with the values utilized in the Monte Carlo simulation of the RPC response and based on an analysis of broadband muons from the 120 GeV primary proton beam together with a 9 feet (3 m) iron beam blocker [7]. The agreement is reasonably good. The lower values of the efficiency in layers 4 and 5 might be due to muons ranging out in the calorimeter. In principle, the product $\varepsilon_i\mu_i$ can be used to correct for deviations of individual layers from the average response [7]. However, since the values are close to the average value used in the simulation of the RPC response, in the following analysis layer-to-layer corrections were not deemed necessary.

After applying the MIP selection, Fig. 4 shows the total number of hits in all six layers of the calorimeter. The distribution is fit to the sum of a Gaussian distribution and a modified exponential

$$\mathbf{y} = \alpha e^{-\frac{1}{2}(\frac{\mathbf{x}-\beta}{\gamma})^2} + \delta(\mathbf{x} - \mathbf{x_0})^\epsilon e^{\phi(\mathbf{x_0}-\mathbf{x})} \tag{1}$$

where x is the number of hits, $x_0$ is the x value of the first non-zero bin and α, β, γ, δ, ε, and φ are free parameters. Given the large number of parameters the fit is naturally able to describe the data adequately.

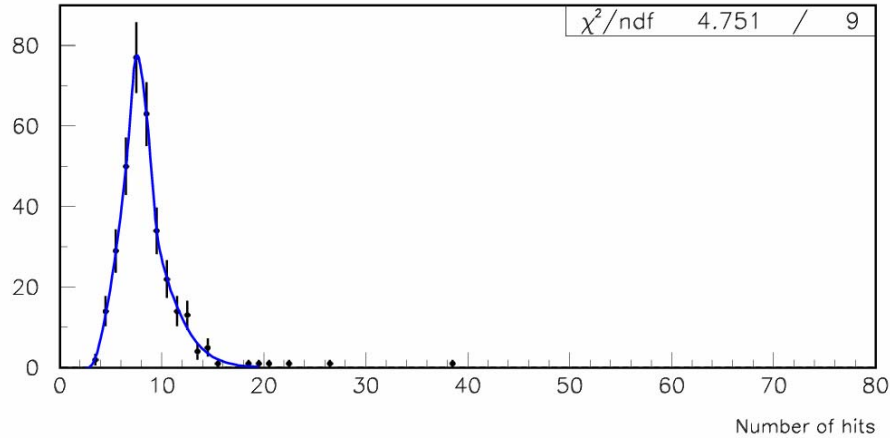

**Figure 4.** Total number of hits in the six layers of the calorimeter for muons obtained with a 2 GeV secondary beam with the stack of iron bricks placed in front of the calorimeter. The blue line is a fit to the sum of a Gaussian and a modified exponential.

As expected, only few events of this data sample passed the shower selection.

## MIP SELECTION

The MIP selection suppresses events with interactions before the third layer of the calorimeter, but still leaves a finite probability for pions to interact in the later layers. Applying this selection, Fig. 5 shows the distribution of the total number of hits in the calorimeter for the various beam settings. The peak around a value of eight originates from muons and pions, which have not interacted in the calorimeter. The tail towards larger values is due to pions (and a small percentage of positrons) which have interacted in the later layers. The distributions are fit to three components:

i) The contribution from pions, obtained from the GEANT4+RPCSIM simulation after applying the same event selection as for the data. The Monte Carlo distributions were empirically fitted to the functional form of Eqn. (1) (1 GeV) or to a Gaussian with variable width [13] (2, 4, 8, and 16 GeV).
ii) The contribution from positrons, also obtained from the GEANT4+RPCSIM simulation of positrons after applying the same event selection. The Monte Carlo distributions were fitted to a Gaussian function.
iii) The contribution from muons, obtained from the fit to the data in Fig.4.

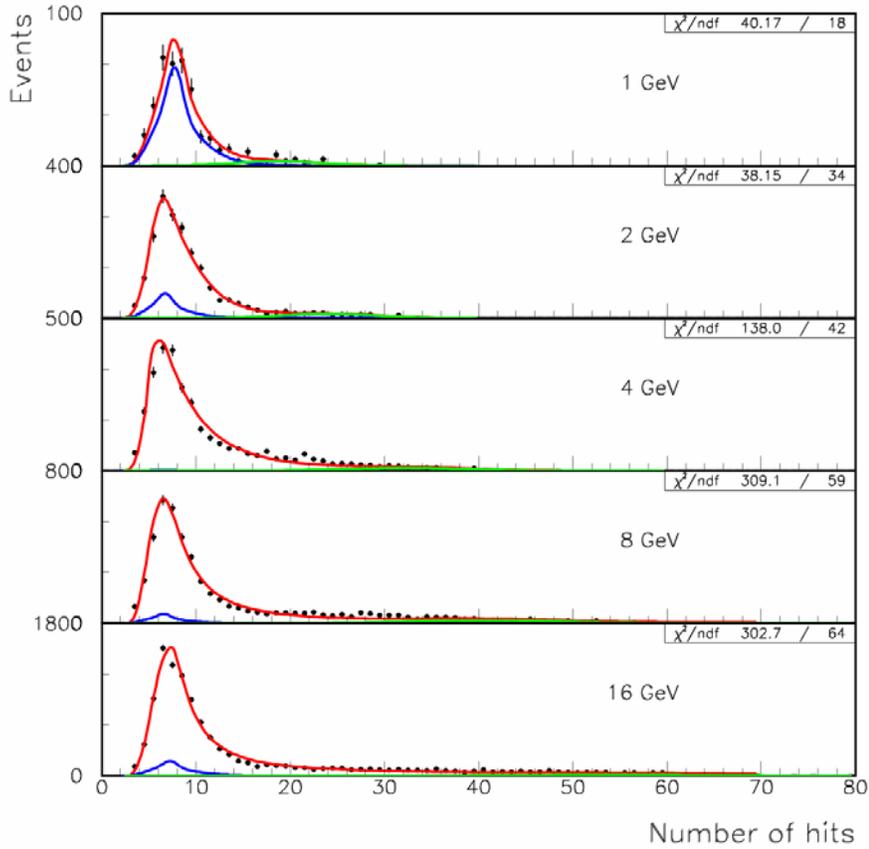

**Figure 5.** MIP selection: Distribution of the number of hits for the various beam energies. The red line shows the result of a fit to three components: pions, positrons and muons. The muon (positron) contributions to the fit are shown in blue (green).

Due to the limited rate capability of RPCs [12], the efficiency around the beam spot decreased, in some layers even up to 30%. This effect is larger than anticipated given the rate of charged particles, the lateral profile and particle composition of the beam, and the predicted number of avalanches in individual layers for positron and pion induced events. This additional loss of efficiency, however, can be explained as originating from a substantial flux of asynchronous photons (compared to charged particles) in the beam line, mostly at the lower beam energies[4]. No attempt was made to incorporate these effects into the simulation. Rather, the overall number of hits in the three contributions of the fit was scaled to reproduce the left slope of the distributions. This constitutes the only adjustment to the simulation of the RPC response, as previously established with the muon [7] and positron [11] data.

The muon content appears to be large at 1 GeV, leaving only a small fraction to pions. With increasing energy the muon content decreases. Overall, the contamination from positrons, which failed to trigger the Čerenkov counters, is quite small. The fits reproduce the general features of the data, but also show some significant deviations, leading to poor $\chi^2$ – values. The shape of the MIP peak was found to depend strongly on the angular distribution of the beam particles, which might not have been reproduced perfectly in the simulation.

## SHOWER SELECTION

With a digital hadron calorimeter the energy of an incoming hadron, $E_{hadron}$, can be reconstructed from the number of hits associated with that particle. Ignoring effects of high density sub-clusters, which might require non-linear corrections, the sum of hits is expected to be proportional to $E_{hadron}$. However, in the present tests, given the limited depth of the calorimeter, the response is not expected to be linear.

Applying the shower selection, Fig. 6 shows the total number of hits in the calorimeter for the various beam energies. The distributions were fit to two components:

i) The contribution from pions, obtained from the GEANT4+RPCSIM simulation and by applying the same event selection as for the data. The Monte Carlo distributions were empirically fitted to the second term only (1, 2, 4, and, 8 GeV) or to both terms of Eqn. (1) (16 GeV).
ii) The contribution from positrons, obtained from the GEANT4+RPCSIM simulation of positrons and by applying the same event selection. The Monte Carlo distributions were fit to a Gaussian function.

Due to the large lateral size of hadronic showers, the effect of the rate limitations of RPCs was not as severe as for the MIP selection. Therefore, apart from the 16 GeV data, no scaling of the Monte Carlo distributions was necessary. Hence, the Monte Carlo simulations may be considered as absolute predictions. The 16 GeV predictions were adjusted by -9%.

At lower beam energies only few events pass the shower selection. Among these a large fraction appears to be originating from the positron contamination of the sample. The fits adequately reproduce the measured distributions.

---

[4] Subsequent measurements with scintillator counters in the MTBF beam line confirmed the presence of a significant flux of asynchronous (to charged particles) photons in the beam line.

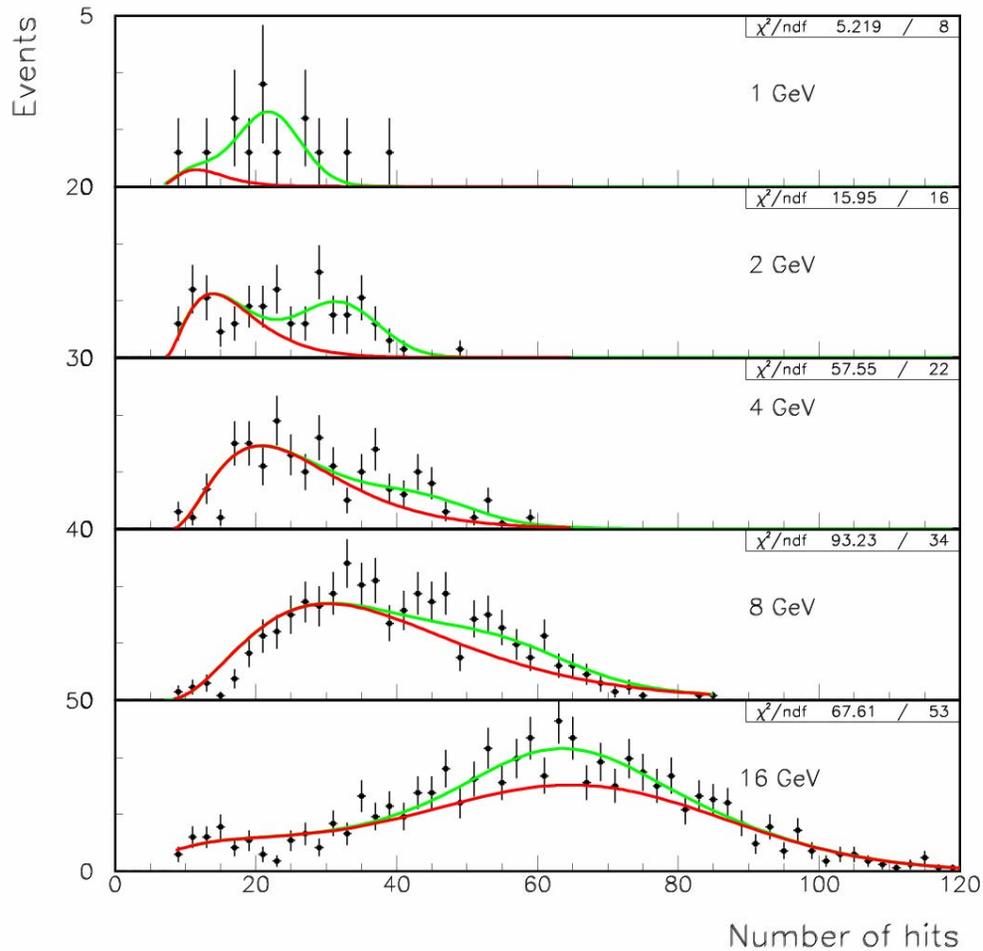

**Figure 6.** Shower selection: Distribution of the total number of hits in the calorimeter. The lines indicate the results of the fits, where the red line shows the pion contribution and the green line represents the positron contribution as added on top of the pion curves.

## PREDICTIONS FOR AN EXTENDED CALORIMETER

Encouraged by the good description of hadronic showers in the present small prototype stack, the simulation tools were used to predict the performance of an extended calorimeter featuring the same readout segmentation, absorber structure and chamber performance. In order to minimize effects due to energy leakage and to study the performance of an RPC-based calorimeter per se, the extended calorimeter included 107 planes, each with an area of 1.5 x 1.5 m$^2$. The depth of the calorimeter thus corresponded to approximately 13 nuclear interaction lengths. The pions were generated such that they entered the calorimeter in the center of the first plane.

Figure 7 shows the distribution of hits in the calorimeter for a selection of pion energies between 1 and 60 GeV. The distributions have been fitted to Gaussian functions, shown as solid lines in the figures.

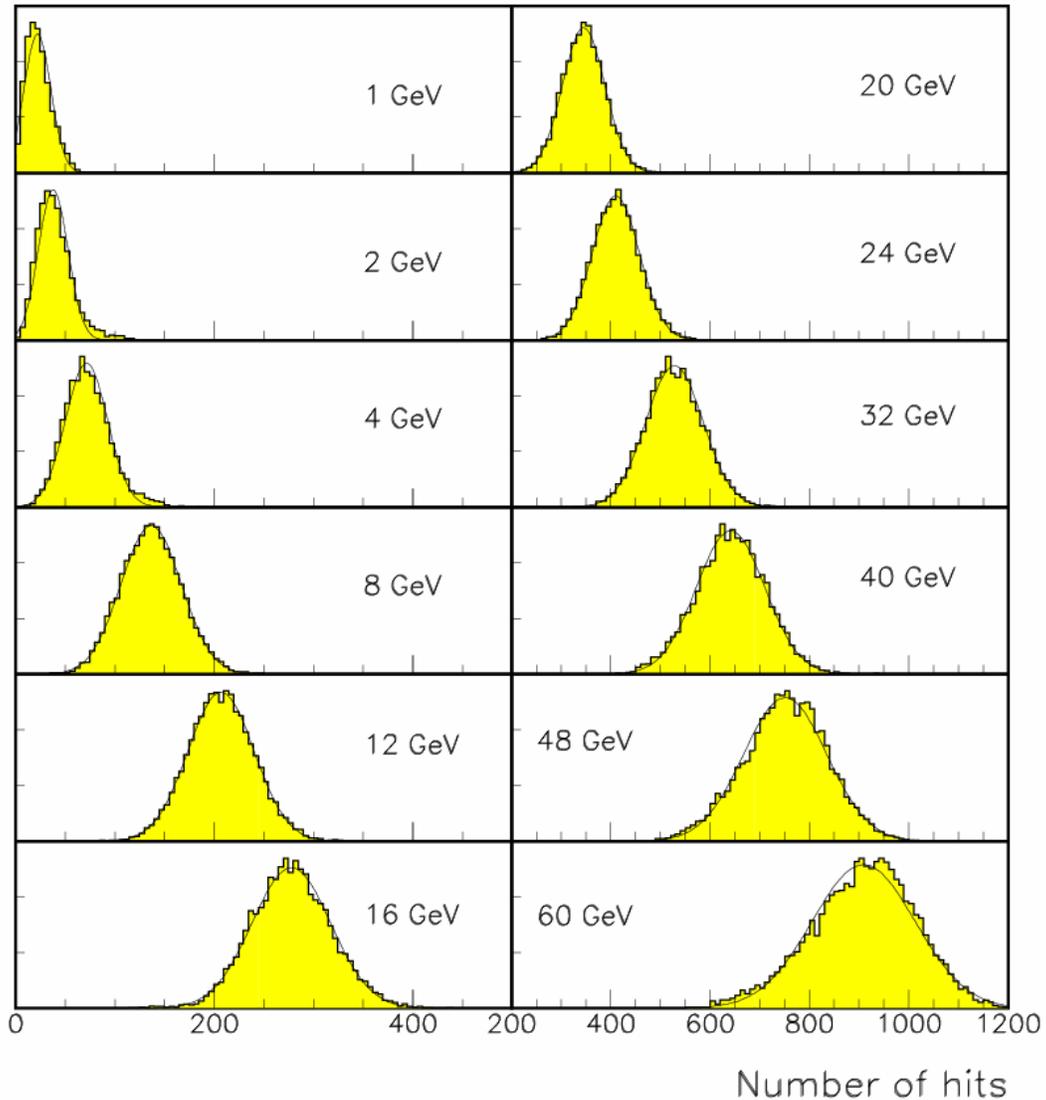

**Figure 7.** Predicted distributions of the number of hits in the extended calorimeter for a selection of pion energies between 1 and 60 GeV. The simulation includes the modeling of hadronic showers (based on GEANT4) and the response of RPC (using the RPCSIM program). The lines are the results of fits to Gaussian distribution functions.

The mean values of the fits versus the energy of the incident pions are shown in the top part of Fig. 8. The means for energies between 1 and 20 GeV have been fitted to a straight line, shown in the figure as solid line. To expose a possible non-linearity of the response for energies above 20 GeV, the fitted line was extended up to 60 GeV

(shown as a dash-dotted line in Fig. 8). At 60 GeV the response is seen to be about 10% smaller than expected from a perfectly linear behavior. This effect is related to the probability of overlap of multiple charged particles in a single readout pad. As expected, studies with smaller pad sizes resulted in a significantly extended range of the linear response.

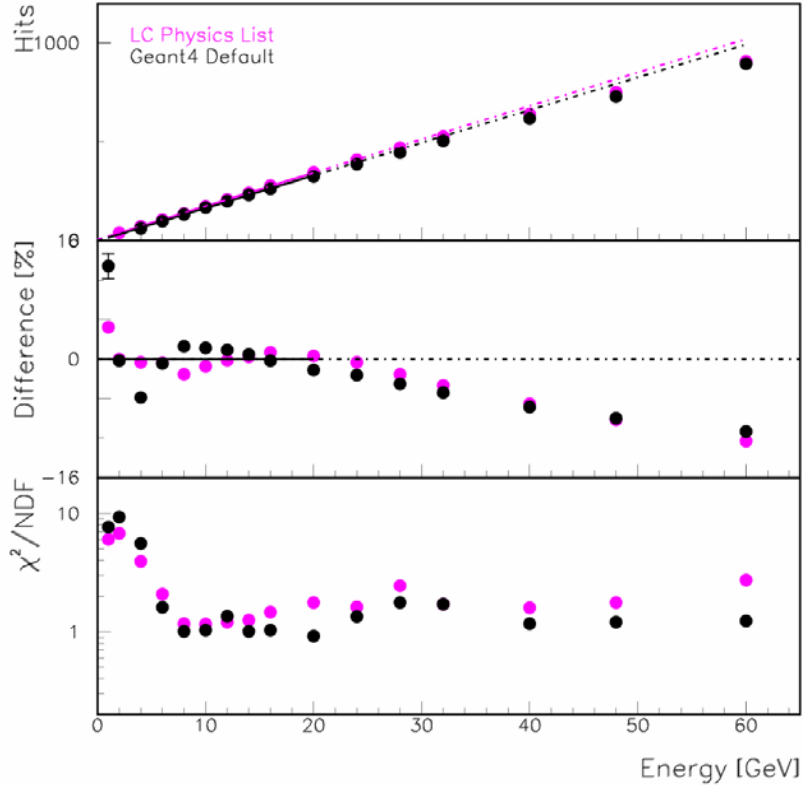

**Figure 8.** Top: Means obtained from Gaussian fits to the distribution of the number of hits as function of pion energy. The solid lines are the results of linear fits to the data points in the range of 1 – 20 GeV. The dash-dotted lines are extrapolations of these lines to higher energies. The simulations are based on the LC Physics list (magenta dots) and the Fast and Simple Physics list (black dots). Middle: Difference between the means and the (extrapolated) lines obtained from the above linear fit. Bottom: $\chi^2$ – values obtained from Gaussian fits to the distributions of the number of hits as function of pion energy.

The middle part of Fig. 8 shows the difference between the linear fit (up to 20 GeV) and the means of the distributions, while the lower part of the figure shows the $\chi^2$ values of the fits. A discontinuity is seen around 8 GeV, which coincides with the transition between the lower energies, where the Gaussian function poorly describes the data and the higher energies where the fits are satisfactory. At lower energies the distributions show large tails towards higher energies. These are due to pions which interact late in the stack and therefore create hits in a large number of layers without undergoing significant energy loss. This effect is also present at higher energies, but appears less noticeable due to the overall large number of hits following the first nuclear interaction.

In order to understand this discontinuity, the simulations were repeated with the Fast and Simple Physics List [9]. The results, also displayed in Fig. 8, show a discontinuity as well, but here centered on a pion energy of 4 GeV. At this moment it is not understood if these discontinuities are due to the modeling of hadronic showers or are an intrinsic feature of an RPC-based DHCAL. Test beam data with a large calorimeter test module, now under construction, will be essential to shed light on this issue.

Figure 9 shows the widths of the Gaussian distributions divided by the corresponding mean of the distribution as a function of pion energy. The data were fitted to the quadratic sum of a stochastic and a constant term:

$$\frac{\sigma}{E} = \frac{\alpha}{\sqrt{E(GeV)}} \oplus \frac{c}{E} \tag{2}$$

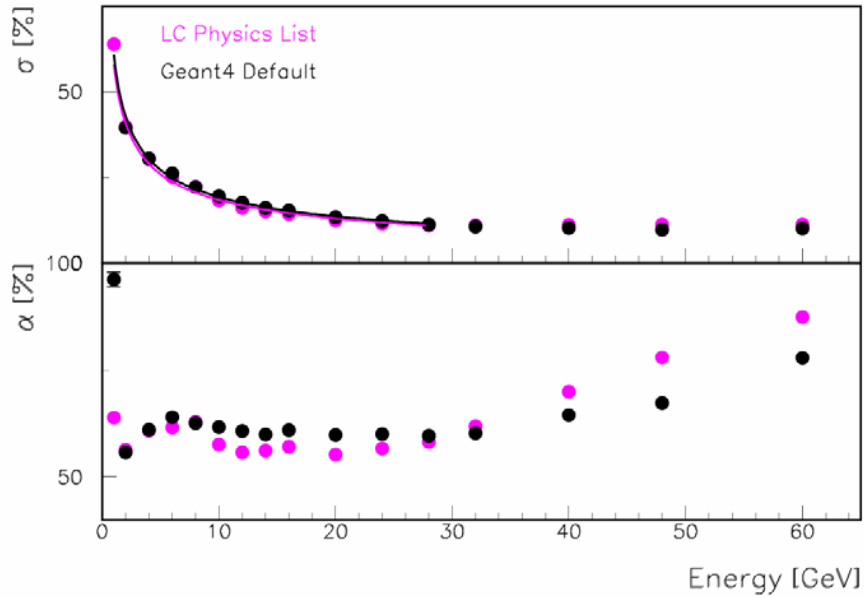

**Figure 9.** Top: Relative width obtained from Gaussian fits to the distribution of the number of hits versus pion energy. The lines are fits to the quadratic sum of a stochastic and a constant term. The simulations are based on the LC Physics list (magenta dots) and the Fast and Simple Physics list (black dots). Bottom: Stochastic terms (for energies up to 20 GeV) or Gaussian widths multiplied by √E for energies above 20 GeV.

The fit yielded a stochastic term of the order of 58% and a negligible constant term. At energies above 20 GeV, the increased probability of multiple particles overlapping in a single pad deteriorates the resolution. This effect is most visible in the lower part of Fig. 9 which shows the stochastic terms (for energies up to 20 GeV) or the Gaussian widths multiplied by √E(GeV) for energies above 20 GeV. Again, a reduction in the readout pad size improves the results at higher energies (not shown).

Finally, the effect of a smaller than 100% MIP detection efficiency and of different average pad multiplicities were investigated. As expected, the calibration (number of pads per GeV) depends strongly on the average efficiency and pad multiplicity. However, variations in the average chamber performance had only a minor effect on the energy resolution.

## CONCLUSIONS

A prototype digital hadron calorimeter (DHCAL) with Resistive Plate Chambers (RPCs) as active elements was exposed to pions in the energy range of 1 – 16 GeV. The calorimeter consisted of six layers interleaved with absorber plates with a thickness corresponding to 1.2 $X_0$. The overall depth of the calorimeter corresponded to about 0.65 nuclear interaction lengths.

The set-up has been simulated by a GEANT4 based program together with a standalone program to model the response of the RPCs (RPCSIM). Three parameters of the response simulation were tuned using data from a broad band muon beam. The last parameter, a short-range distance cut for the efficiency of RPCs, was tuned using positron data.

Measurements of the response of the calorimeter have been presented, for both a 'MIP selection' and a 'shower selection'. In general, the simulation reproduces the data quite well. However, some significant deficiencies of hits in the MIP selection data are observed, mostly in the high rate regions of the calorimeter. This effect amounts up to 30% in some regions of some layers and is understood as being due to a loss of efficiency related to high particle fluxes in these regions.

Predictions for an extended RPC-based DHCAL show a linear behavior up to pion energies of 20 GeV and a 10% non-linearity at 60 GeV. Fits to the widths of the response up to 20 GeV yield a stochastic term of ~58% and a negligible constant term.

## ACKNOWLEDGEMENTS

We would like to thank the Fermilab test beam crew, in particular Erik Ramberg, Doug Jensen, Rick Coleman and Chuck Brown, for providing us with excellent beam. The University of Texas at Arlington ILC group is acknowledged for providing the two trigger scintillator paddles and their associated trigger logic.